\documentclass[aps,prd,reprint,groupedaddress,nofootinbib]{revtex4-2}
\usepackage{graphicx}
\input{epsf}
\begin{document}

\title{ 
\bf Contribution of hard photon emission to charge asymmetry in elastic (anti)lepton-proton scattering
}
\author{ A. Afanasev}
\email[]{afanas@gwu.edu}
\affiliation{
Department of Physics, the
George Washington University,
Washington, DC 20052 USA}
\author{ A. Ilyichev}
\email[]{ily@hep.by}
\affiliation{
Institute for Nuclear Problems,
 Belarusian State University,
220040  Minsk,  Belarus}
\date{\today}
\begin{abstract}
The influence of the hard photon emission on the charge asymmetry in the (anti)lepton-proton elastic scattering was evaluated for the first  time beyond the ultrarelativistic limit, while retaining the lepton mass at all steps of the calculation. This contribution - responsible for the charge asymmetry - is induced 
by interference between real photon emission from the lepton and proton. During the calculation any excited states of
the intermediated proton are not considered, allowing us to use the standard fermionic propagator for this particle. 
The infrared divergence extracted using Lorenz-invariant approach of Bardin-Shumeiko is canceled by the corresponding soft part of the two-photon exchange contribution. Numerical analysis was performed  within kinematic conditions of Jefferson Lab measurements and MUSE experiment in PSI.
\end{abstract}

\maketitle
\section{Introduction}
The elastic form factors of the proton play an essential role in our understanding the nucleon electromagnetic
structure. However $Q^2$-dependence of the ratio of the electric to magnetic proton form factors, $G_E(Q^2)/G_M(Q^2)$,
obtained from the unpolarized \cite{Andivahis94,Qattan05} and polarized \cite{Jones,Gayou} electron elastic scattering data disagreed with each other, and the disagreement was attributed to two-photon exchange effects. The challenge in computing two-photon exchange contribution is in the need for modeling nucleon's structure. It motivated a dedicated program of both direct and indirect experimental measurements of two-photon effects in electron scattering, along with a significant theory effort; the recent status of the problem is reviewed in, $e.g.$, Ref.\cite{Afanasev_review}. Moreover, two-photon effects may have an impact on determination of proton's radius and resolution of the ``proton radius puzzle" \cite{Pohl}. Direct measurements of the two-photon effect on the unpolarized scattering cross section \cite{Jlab,Jlab1,Rachek15,Henderson17} are based on the observation that this contribution changes its sign with a sign of the scattering lepton, therefore it can be evaluated via measurements of  the charge asymmetry. However, the charge asymmetry is caused not only by the two-photon exchange, but also - as the requirement of the infrared divergence cancellation - by interference of real photon emission from the lepton and proton, which is a subject of this paper.

Current research efforts aimed at resolving the above-mentioned ``proton radius puzzle'' \cite{Pohl} include, in particular, comparison -- with a sub-percent accuracy -- of muon/anti-muon and electron/positron scattering on a proton target in an ongoing MUSE experiment \cite{MUSE} in PSI. In MUSE, the muon momenta are in 100-200 MeV/c range, of the order of muon's mass, requiring that both the leading-order expressions for the scattering cross section and QED corrections include muon's mass with no ultra-relativistic approximations usually applied for scattering of electrons.


In many cases for the estimation of the high order QED effects to the exclusive processes
the loop corrections (with the additional particle contributions) are calculated exactly or within ultrarelativistic approximation (with respect to lepton's mass)
while the real photon emission is considered within the soft photon approximation. 
For example, in the papers \cite{Kaiser} and \cite{Vanderhaeghen} 
for M\"oller and virtual Compton scattering processes, respectively,
the virtual QED corrections have been calculated beyond
the ultrarelativistic limit but only the soft part of the real photon emission was  taken into account.

The charge asymmetry in the leading order of QED in the soft-photon approximation was calculated in Ref.\cite{Koshchii}, while treating the lepton mass exactly. In the present paper we go beyond the soft-photon approximation and include, for the first time, the effects of hard-photon emission and  its influence on the charge asymmetry in elastic (anti)lepton-proton scattering. We study effects of hard-photon emission in various kinematic conditions of experiments targeting the charge asymmetry: MeV energies of MUSE \cite{MUSE} and GeV energies of JLab \cite{Jlab, Jlab1}.  The infrared divergence is  canceled with the corresponding soft part from the two-photon exchange using Lorentz-invariant approach of Bardin-Shumeiko \cite{BSh}.  All calculations were performed  without using an ultra-relativistic limit, making them applicable for both low and high energies of scattering leptons.

\section{Method of calculation}
During our calculation we assume that there is no excited states of the intermediated proton. As a result, the proton
propagator looks like a standard fermionic one. The second assumption is that the on-shell proton vertex, 
\begin{eqnarray}
\Gamma_\mu (q)=\gamma_\mu F_d(-q^2)+\frac {i\sigma_{\mu \nu}q^\nu}{2M}F_p(-q^2),
\end{eqnarray}
is applicable  within off-shell region. Here $\sigma_{\mu \nu}=i[\gamma_\mu,\gamma_\nu]/2$,
$M$ is a proton mass, $q$ is a four-momentum of the  virtual photon.
The Dirac and Pauli form factors can be expressed through the electromagnetic ones:    
\begin{eqnarray}
F_d(-q^2)&=&\frac{G_E(-q^2)+\tau G_M(-q^2)}{1+\tau},
\nonumber\\
F_p(-q^2)&=&\frac{G_M(-q^2)-G_E(-q^2)}{1+\tau},
\end{eqnarray}
where $\tau=-q^2/4M^2$.

The lowest-order (Born) contribution to the elastic $l^\mp p$ scattering is
presented by Feynman graphs in Fig.~\ref{fig1} and it can be described by the following matrix elements:  
\begin{figure}

\includegraphics[width=3cm,height=4cm]{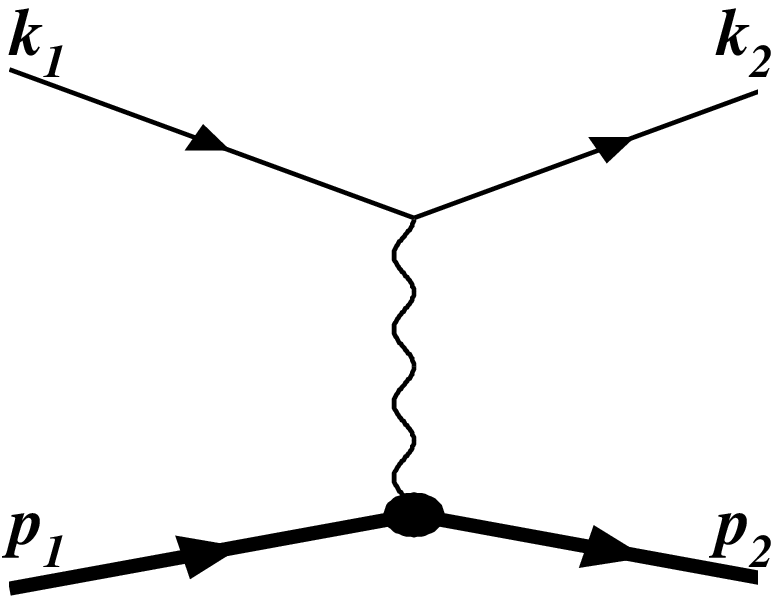}
\hspace*{5mm}
\includegraphics[width=3cm,height=4cm]{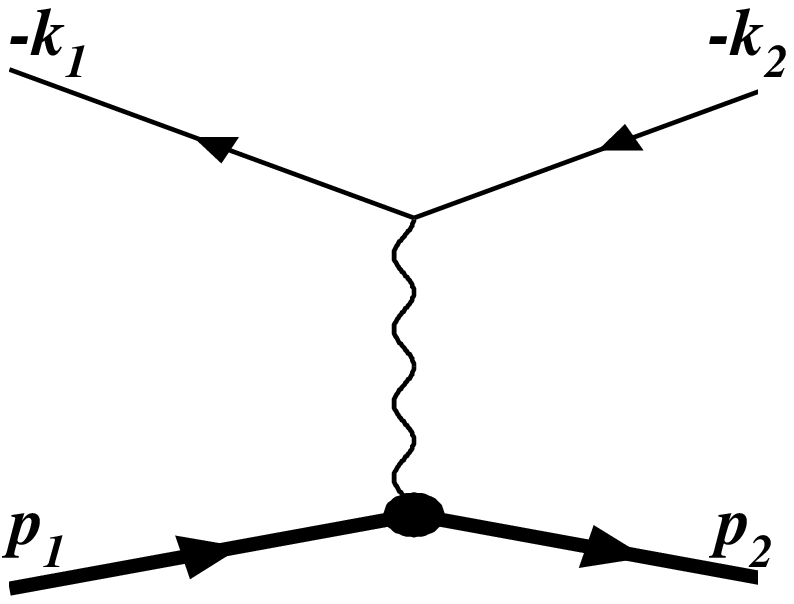}
\put(-199,10){\mbox{a)}}
\put(-60,10){\mbox{b)}}
\caption{Feynman graphs for the lowest-order
contribution to elastic $l^-p$ (a) and $l^+p$ (b) scattering.
}
\label{fig1}
\end{figure}
\begin{eqnarray}
{\cal M}^-_b&=&\frac{ie^2}{Q^2}
{\bar u}(k_2)\gamma^\mu u(k_1)
{\bar U}(p_2)\Gamma_\mu (q)U(p_1),
\nonumber\\
{\cal M}^+_b&=&\frac{ie^2}{Q^2}
{\bar u}(-k_1)\gamma^\mu u(-k_2)
{\bar U}(p_2)\Gamma_\mu (q)U(p_1),
\end{eqnarray}
where $Q^2=-q^2=-(k_1-k_2)^2$ and $e=\sqrt{4\pi\alpha}$.
Since the squares of these two matrix elements are identical and are insensitive to the sign of lepton's charge, it is not possible to distinguish
the lepton-proton from antilepton-proton scattering processes at the one-photon exchange
level. Their contribution 
to the cross section can be written as
\begin{eqnarray}
d\sigma_B&=&\frac 1{2\sqrt{\lambda_S}}|{\cal M}^\mp_b|^2 d\Gamma_2, 
\end{eqnarray}
where $\lambda_S=S^2-4m^2M^2$, $S=2p_1k_1$, $m$ is a lepton mass and the phase space reads:  
\begin{eqnarray}
d\Gamma_2&=&(2\pi)^4\delta ^4(p_1+k_1-p_2-k_2)
\frac{d^3k_2}{(2\pi)^32k_{20}}
\frac{d^3p_2}{(2\pi)^32p_{20}}
\nonumber\\
&=&\frac{dQ^2}{8\pi\sqrt{\lambda_S}}.
\end{eqnarray}

\begin{figure}[t]
\includegraphics[width=3cm,height=4cm]{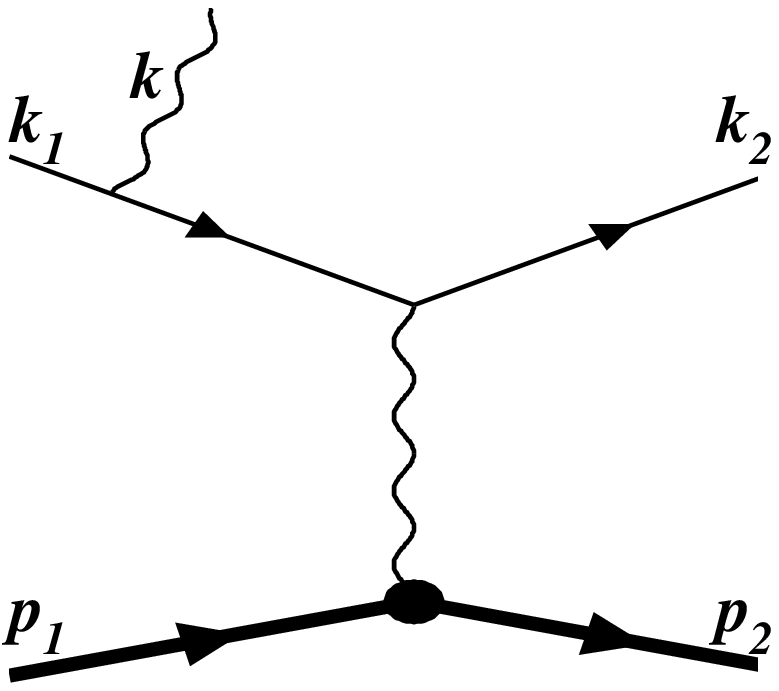}
\includegraphics[width=3cm,height=4cm]{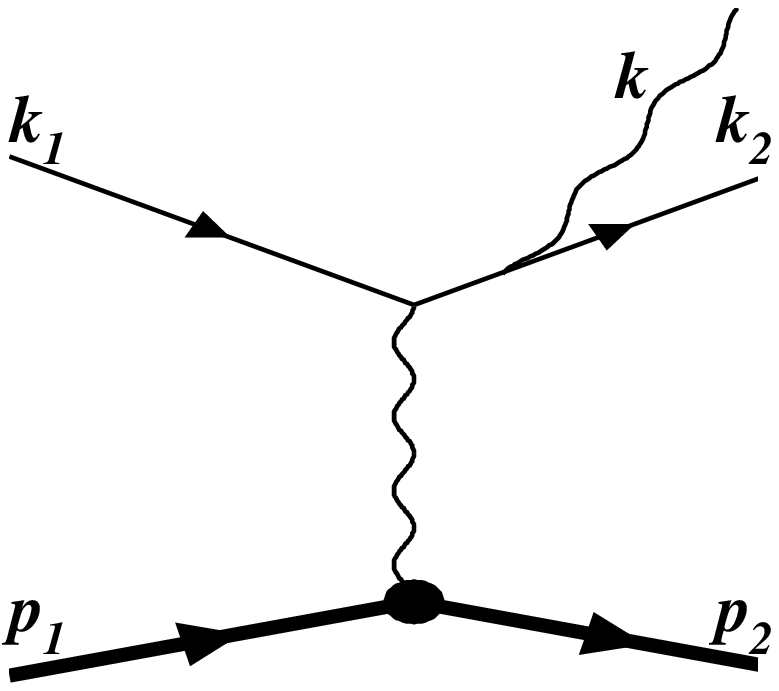}
\put(-135,10){\mbox{a)}}
\put(-47,10){\mbox{b)}}
\\[-6mm]
\includegraphics[width=3cm,height=4cm]{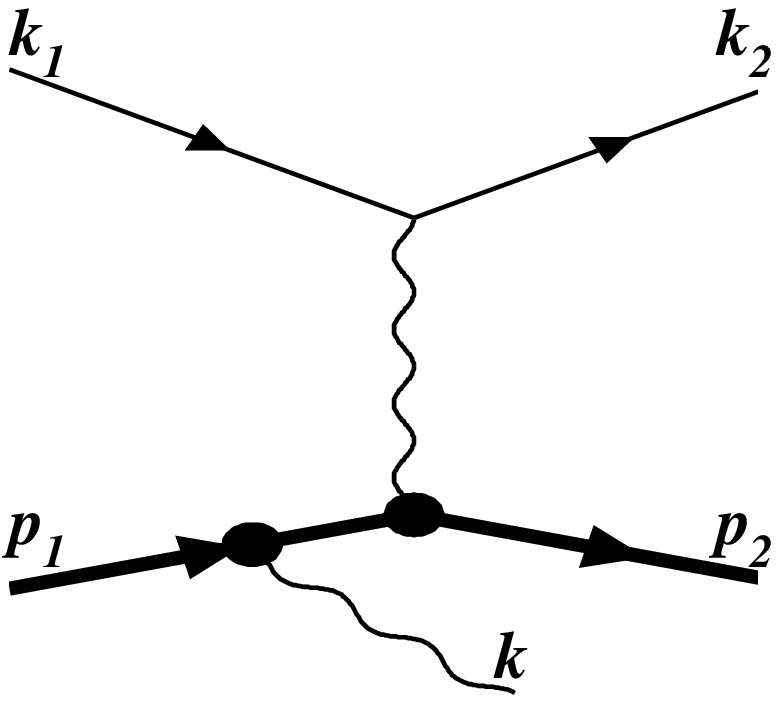}
\includegraphics[width=3cm,height=4cm]{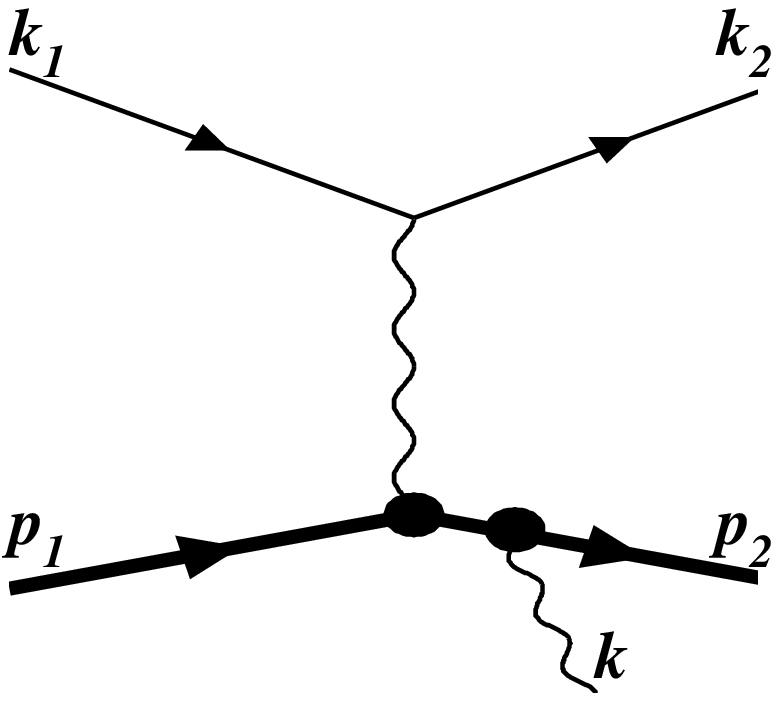}
\put(-135,0){\mbox{c)}}
\put(-47,0){\mbox{d)}}
\\[-6mm]
\includegraphics[width=3cm,height=4cm]{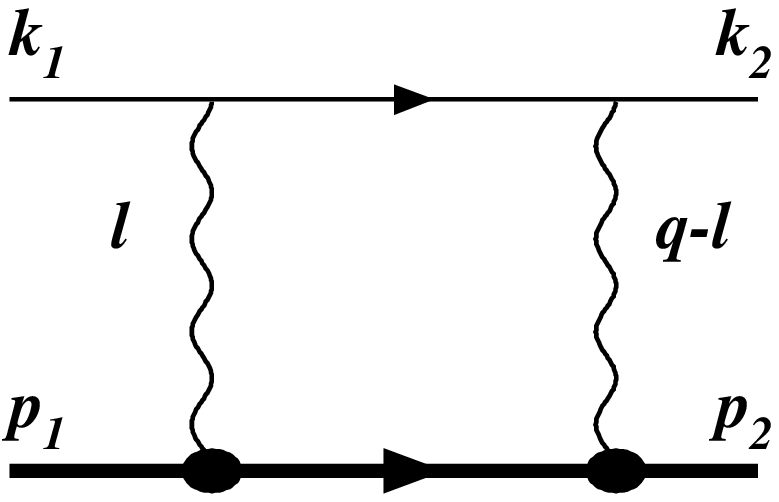}
\includegraphics[width=3cm,height=4cm]{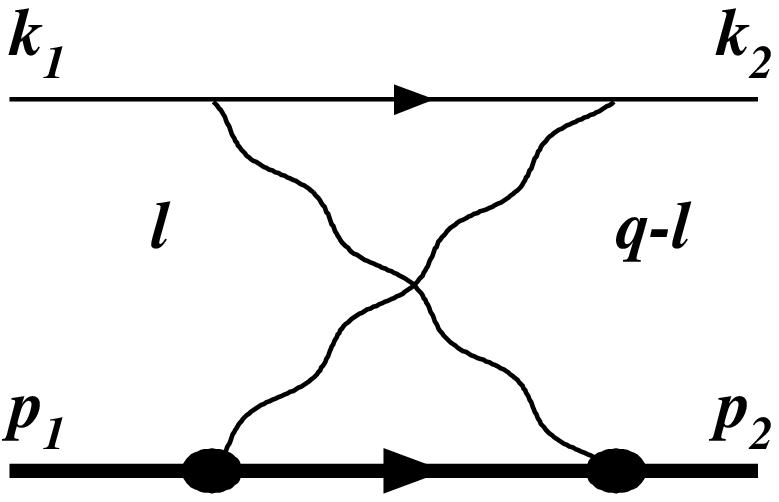}
\put(-135,0){\mbox{e)}}
\put(-47,0){\mbox{f)}}
\caption{Feynman graphs for the real photon emission from the lepton (a,b) and proton (c,d) legs as well as the direct (e) and
cross (f) two-photon exchange within $l^-p$-scattering. The similar graphs for $l^+p$ scattering processes have
an opposite direction  for the leptonic arrows and a negative sign for its momenta.
}
\label{fig2}
\end{figure}

The Feynman graphs with
the real photon emission both from the lepton and proton
legs are shown in Fig~\ref{fig2} (a-d) for $l^-p$ scattering.
The matrix elements corresponding to these processes as well as 
the real photon emission in $l^+p$ scattering read:
\begin{eqnarray}
{\cal M}^{-}_{lR}
&=-&\frac{ie^3}t
{\bar u}(k_2) \varepsilon_\alpha \Gamma_{lR}^{\mu \alpha} u(k_1)
{\bar U}(p_2)\Gamma_\mu(q-k)U(p_1),
\nonumber\\
{\cal M}^{+}_{lR}
&=&\frac{ie^3}t
{\bar u}(-k_1) \varepsilon_\alpha {\bar \Gamma}_{lR}^{\mu \alpha} u(-k_2)
{\bar U}(p_2)\Gamma_\mu(q-k)U(p_1),
\nonumber\\
{\cal M}^{-}_{hR}
&=&\frac{ie^3}{Q^2}{\bar u}(k_2)\gamma_\mu u(k_1){\bar U}(p_2)\varepsilon_\alpha \Gamma_{hR}^{\mu \alpha}U(p_1),
\nonumber\\
{\cal M}^{+}_{hR}
&=&\frac{ie^3}{Q^2}{\bar u}(-k_1)\gamma_\mu u(-k_2){\bar U}(p_2)\varepsilon_\alpha \Gamma_{hR}^{\mu \alpha}U(p_1),
\label{mlhr}
\end{eqnarray}
where $t=-(q-k)^2=-(p_2-p_1)^2$, $\varepsilon_\alpha$ is the photon polarized vector and
\begin{eqnarray}
\Gamma_{lR}^{\mu \alpha}&=&\Biggl(
\frac{k_{1\alpha }}{kk_1}- 
\frac{k_{2\alpha }}{kk_2}\Biggr)\gamma^\mu
-\frac{\gamma^\mu \hat k \gamma^\alpha}{2k_1k} 
-\frac{ \gamma^\alpha\hat k \gamma^\mu}{2k_2k},
\nonumber\\
{\bar \Gamma}_{lR}^{\mu \alpha}&=&\Biggl(
\frac{k_{1\alpha }}{kk_1}- 
\frac{k_{2\alpha }}{kk_2}\Biggr)\gamma^\mu
-\frac{ \gamma^\alpha\hat k \gamma^\mu}{2k_1k}
-\frac{\gamma^\mu \hat k \gamma^\alpha}{2k_2k}, 
\nonumber\\
\Gamma_{hR}^{\mu \alpha}&=&
\Gamma^\mu(q)\frac{\hat p_1-\hat k+M}{2p_1k}\Gamma^\alpha (-k)
\nonumber\\&&
-\Gamma^\alpha (-k) 
\frac{\hat p_2+\hat k+M}{2p_2k}\Gamma^\mu(q).
\label{vr}
\end{eqnarray}

The part of the cross section with 
the interference between the real photon emissions from hadron and lepton lines reads:
\begin{eqnarray}
d\sigma_R^\mp=\frac 1{2\sqrt{\lambda_S}}({\cal M}^\mp_{lR}{\cal M}^{\mp\;\dagger}_{hR}+{\cal M}^\mp_{hR}{\cal M}^{\mp\;\dagger}_{lR})
d\Gamma_3,
\label{sri}
\end{eqnarray}
where the phase space has a form:
\begin{eqnarray}
d\Gamma_3&=&(2\pi)^4\delta ^4(p_1+k_1-p_2-k_2-k)
\nonumber\\&&\times
\frac{d^3k}{(2\pi)^32k_{0}}
\frac{d^3k_2}{(2\pi)^32k_{20}}
\frac{d^3p_2}{(2\pi)^32p_{20}}
\nonumber\\
&=&\frac{dQ^2dv dtd\phi_k}{2^8\pi^4\sqrt{\lambda_S Q^2(Q^2+4M^2)}}.
\end{eqnarray}
Here $v=(p_1+q)^2-M^2$ is a photonic variable defining inelasticity, $\phi_k$ is an angle between
(${\bf k}_1$,${\bf k}_2$) and (${\bf k}$,${\bf q}$) planes in the rest frame (${\bf p}_1=0$).
Performing the direct calculation, one can find that the interference terms for $l^-p$ and $l^+p$ 
have opposite signs,
\begin{eqnarray}
d\sigma_R^+=-d\sigma_R^-.
\end{eqnarray}

In order to estimate this contribution to the elastic process, it is  necessary to integrate
$d\sigma_R^\mp$ over three photonic variables: $v$, $t$ and $\phi_k$. 
However, since the expressions~(\ref{sri}) contain the infrared divergent terms  
at $k\to 0$, it is not possible to perform the integration in the straightforward way. 
Following to the Bardin-Shumeiko approach \cite{BSh} for the extraction of the infrared (IR) divergence,
the identity transformation has to be performed:
\begin{eqnarray}
d\sigma_R^\mp=d\sigma_R^\mp-d\sigma_{IR}^\mp+
d\sigma_{IR}^\mp=d\sigma_F^\mp+d\sigma_{IR}^\mp.
\end{eqnarray}

In the infrared-free term $d\sigma_F^\pm$ the integration
can be performed over three photonic variables without any restrictions. 
The infrared term $d\sigma_{IR}^\pm$ can be obtained by the substitution 
to the expressions (\ref{mlhr}) the soft parts of the vertexes with a real photon emission (\ref{vr})
that survive when $k\to 0$: 
\begin{eqnarray}
\Gamma_{lR\;soft}^{\mu \alpha}&=&\Biggl(
\frac{k_{1\alpha }}{kk_1}- 
\frac{k_{2\alpha }}{kk_2}\Biggr)\gamma^\mu
,
\nonumber\\
{\bar \Gamma}_{lR\;soft}^{\mu \alpha}&=&\Biggl(
\frac{k_{1\alpha }}{kk_1}- 
\frac{k_{2\alpha }}{kk_2}\Biggr)\gamma^\mu, 
\nonumber\\
\Gamma_{hR\;soft}^{\mu \alpha}&=&\Biggl(
\frac{p_{1\alpha }}{kp_1}- 
\frac{p_{2\alpha }}{kp_2}\Biggr)
\Gamma^\mu(q).
\label{vrs}
\end{eqnarray}

As a result $d\sigma_{IR}^\pm$ is factorized in front of the Born contribution  
the following way:
\begin{eqnarray}
\frac {d\sigma_{IR}^\mp}{dQ^2}=\mp \frac \alpha \pi (\delta_S+\delta_H)\frac {d\sigma_{B}}{dQ^2}.
\label{irr}
\end{eqnarray}
The quantities $\delta_S$ and $\delta_H$ appear after splitting 
the integration region over inelasticity $v$ by the introduction of the infinitesimal parameter $\bar v$
\begin{eqnarray}
\delta_S=\frac 1\pi \int\limits^{\bar v}_0dv\int \frac{d^3k}{k_0}F_{IR}\delta((p_1+q-k)^2-M^2),
\nonumber\\
\delta_H=\frac 1\pi \int\limits_{\bar v}^{v_m}dv\int \frac{d^3k}{k_0}F_{IR}\delta((p_1+q-k)^2-M^2),
\end{eqnarray}
where
\begin{eqnarray}
F_{IR}&=&-\frac 12
\Biggl(\frac{k_1}{k_1k}-\frac{k_2}{k_2k}\Biggl)
\Biggl(\frac{p_1}{p_1k}-\frac{p_2}{p_2k}\Biggl).
\end{eqnarray}

The upper limits of integration with respect to the variable $v$ is defined as
\begin{eqnarray}
v_m=\frac 1{2m^2}(\sqrt{\lambda_S}\sqrt{Q^2(Q^2+4m^2)}-2m^2Q^2-Q^2S).
\nonumber\\
\end{eqnarray}
In practice, however, the influence of the hard real photon emission 
to the asymmetry can be essentially reduced by applying a cut $v_{cut}$ on the inelasticity which is also a measured
quantity in the elastic lepton-proton scattering.

Performing the integration in $\delta_S$ (using the photon mass $\lambda$ for the infrared divergence
regularization) and $\delta_H$ we can find that     
\begin{eqnarray}
\delta_S&=&\delta_S^1-2(SL_S-XL_{X})\log\left[\frac {\bar v}{M\lambda}\right],
\nonumber\\
\delta_H&=&\delta_H^1-2(SL_S-XL_{X})\log\left[\frac {v_{m}} {\bar v}\right],
\end{eqnarray}
where
\begin{eqnarray}
L_S&=&\frac 1{\sqrt{\lambda_S}}\log\frac {S+\sqrt{\lambda_S}} {S-\sqrt{\lambda_S}}, 
\nonumber\\
L_{X}&=&\frac 1{\sqrt{\lambda_X}}\log\frac {X+\sqrt{\lambda_X}} {X-\sqrt{\lambda_X}},
\end{eqnarray}
$X=S-Q^2$ and $\lambda_X=X^2-4M^2m^2$. The quantities $\delta_S^1$ and $\delta_H^1$ have a rather complicated structure and
depend neither on ${\bar v}$ nor on $\lambda$. As a result, the sum of 
$\delta_S$ and $\delta_H$ is free from the separated parameter ${\bar v}$ but it contains 
dependence on the fictitious photon mass $\lambda$.

As shown in Fig.~\ref{fig2} (e,f), the matrix elements with
two-photon exchange contribution to the elastic $l^\mp p$ scattering can be
separated into the direct ${\cal M}^\mp_{d2\gamma }$ and cross ${\cal M}^\mp_{x2\gamma }$ terms
that can be presented through the loop integration in a following way:

\begin{widetext}
\begin{eqnarray}
{\cal M}^-_{d2\gamma }&=&\frac{e^4}{(2\pi)^4}\int\frac{d^4l}{l^2(l-q)^2}
{\bar u}(k_2)\gamma^\nu\frac{\hat k_1-\hat l+m}{l^2-2k_1l}\gamma^\mu u(k_1)
{\bar U}(p_2)\Gamma_\nu (q-l)\frac{\hat p_1+\hat l+M}{l^2+2p_1l}\Gamma_\mu(l)U(p_1),
\nonumber\\
{\cal M}^+_{d2\gamma }&=&\frac{e^4}{(2\pi)^4}\int\frac{d^4l}{l^2(l-q)^2}
{\bar u}(-k_1)\gamma^\mu\frac{\hat l-\hat k_1+m}{l^2-2k_1l}\gamma^\nu u(-k_2)
{\bar U}(p_2)\Gamma_\nu (q-l)\frac{\hat p_1+\hat l+M}{l^2+2p_1l}\Gamma_\mu(l)U(p_1),
\nonumber\\
{\cal M}^-_{x2\gamma }&=&\frac{e^4}{(2\pi)^4}\int\frac{d^4l}{l^2(l-q)^2}
{\bar u}(k_2)\gamma^\nu\frac{\hat k_1-\hat l+m}{l^2-2k_1l}\gamma^\mu u(k_1)
{\bar U}(p_2)\Gamma_\mu (l)\frac{\hat p_2-\hat l+M}{l^2-2p_2l}\Gamma_\nu(q-l)U(p_1),
\nonumber\\
{\cal M}^+_{x2\gamma }&=&\frac{e^4}{(2\pi)^4}\int\frac{d^4l}{l^2(l-q)^2}
{\bar u}(-k_1)\gamma^\mu\frac{\hat l-\hat k_1+m}{l^2-2k_1l}\gamma^\nu u(-k_2)
{\bar U}(p_2)\Gamma_\mu (l)\frac{\hat p_2-\hat l+M}{l^2-2p_2l}\Gamma_\nu(q-l)U(p_1).
\end{eqnarray}
\end{widetext}
Note that all of these matrix elements contain the infrared divergence at $l=0$ and $l=q$ points.

The lowest-order two-photon exchange contribution to the elastic $l^\mp p$ cross section has a form
\begin{eqnarray}
d\sigma_{2\gamma }^\mp&=&\frac 1{2\sqrt{\lambda_S}}[
{\cal M}^\mp_b 
({\cal M}^\mp_{d2\gamma }+{\cal M}^\mp_{x2\gamma })^\dagger 
\nonumber\\&&
+({\cal M}^\mp_{d2\gamma }+{\cal M}^\mp_{x2\gamma })
{\cal M}^{\mp\; \dagger}_b ] 
d\Gamma_2. 
\end{eqnarray}
Once again we can find that
\begin{eqnarray}
d\sigma_{2\gamma }^+=-d\sigma_{2\gamma }^-.
\end{eqnarray}

The infrared divergence extracted from two-photon exchange contribution  reads 
\begin{eqnarray}
\frac{d\sigma_{2\gamma \; IR}^\mp}{dQ^2}&=&\mp\frac{\alpha }{\pi}\left (\delta _{2\gamma}^1
+(SL_S-XL_{X})\log\left[\frac {Q^2}{\lambda^2}\right]\right)\frac{d\sigma_{B}}{dQ^2},
\nonumber\\&&
\label{irb}
\end{eqnarray}
where the quantity $\delta _{2\gamma}^1$ has a rather complicated structure and
does not depend on the photon mass $\lambda$.

The sum Eq.~(\ref{irr}) with Eq.~(\ref{irb}) 
\begin{eqnarray}
\frac{d\sigma^{\mp}_{IR}}{dQ^2}+\frac{d\sigma^{\mp}_{2\gamma \; IR}}{dQ^2}
=
\mp \frac \alpha \pi 
\delta_{VR}(Q^2)\frac{d\sigma_{B}}{dQ^2}
\end{eqnarray}
is infrared free since 
\begin{eqnarray}
\delta_{VR}(Q^2)
&=&
\Biggl(
(SL_S-XL_{X})\log\left[\frac {Q^2M^2}{v_m^2}\right]
\nonumber\\&&
+\delta_S^1+\delta_H^1
+\delta _{2\gamma}^1
\Biggr)
\end{eqnarray}
does not depend on $\lambda$.

A physical requirement coming from vanishing asymmetry at $Q^2\to 0$
can be provided by the difference $\delta_{VR}(Q^2)$ and its value at $Q^2=0$:
$\hat\delta_{VR}=\delta_{VR}(Q^2)-\delta_{VR}(0)$.

Finally, the lowest order of the charge-odd contribution
to the elastic lepton-proton cross section reads: 
\begin{eqnarray}
\frac{d\sigma^{\mp}_{odd}}{dQ^2}=\frac{d\sigma^{\mp}_F}{dQ^2}
\mp \frac \alpha \pi 
\hat \delta_{VR}\frac{d\sigma_{B}}{dQ^2}.
\end{eqnarray}

\begin{figure*}[hbt]\centering
\includegraphics[width=70mm,height=70mm]{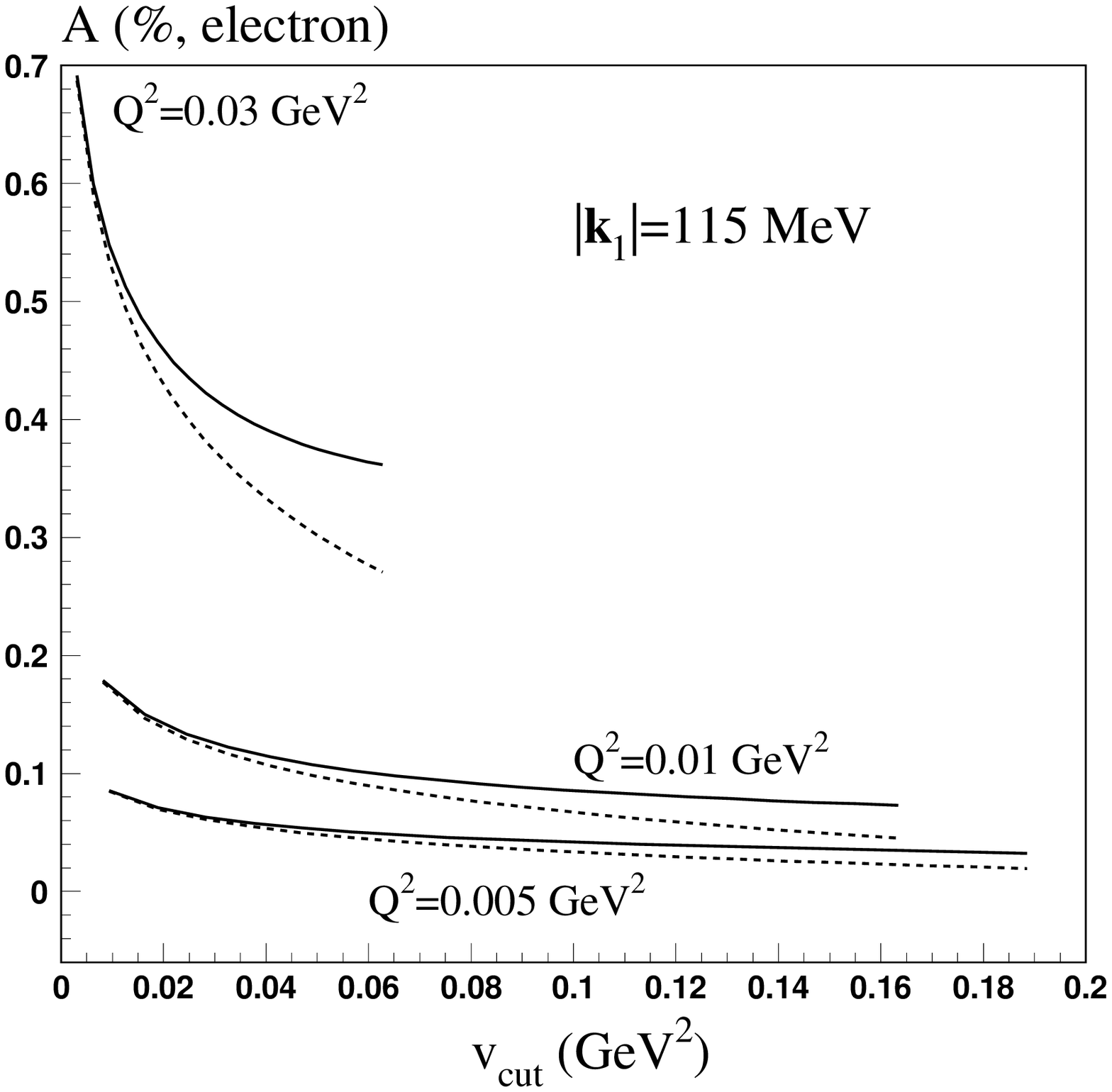}
\hspace*{-6mm}
\includegraphics[width=70mm,height=70mm]{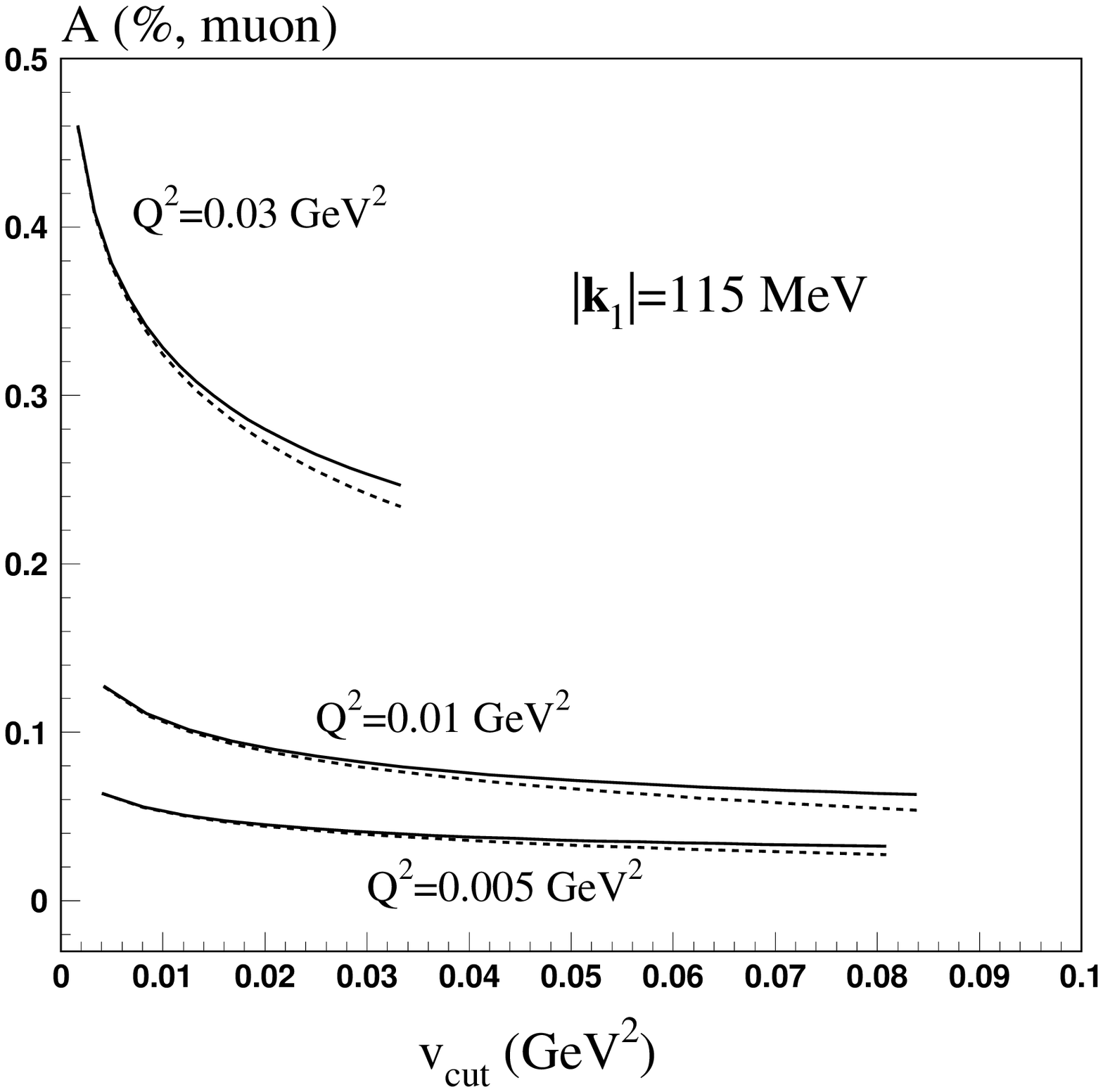}
\\[-9mm]
\includegraphics[width=70mm,height=70mm]{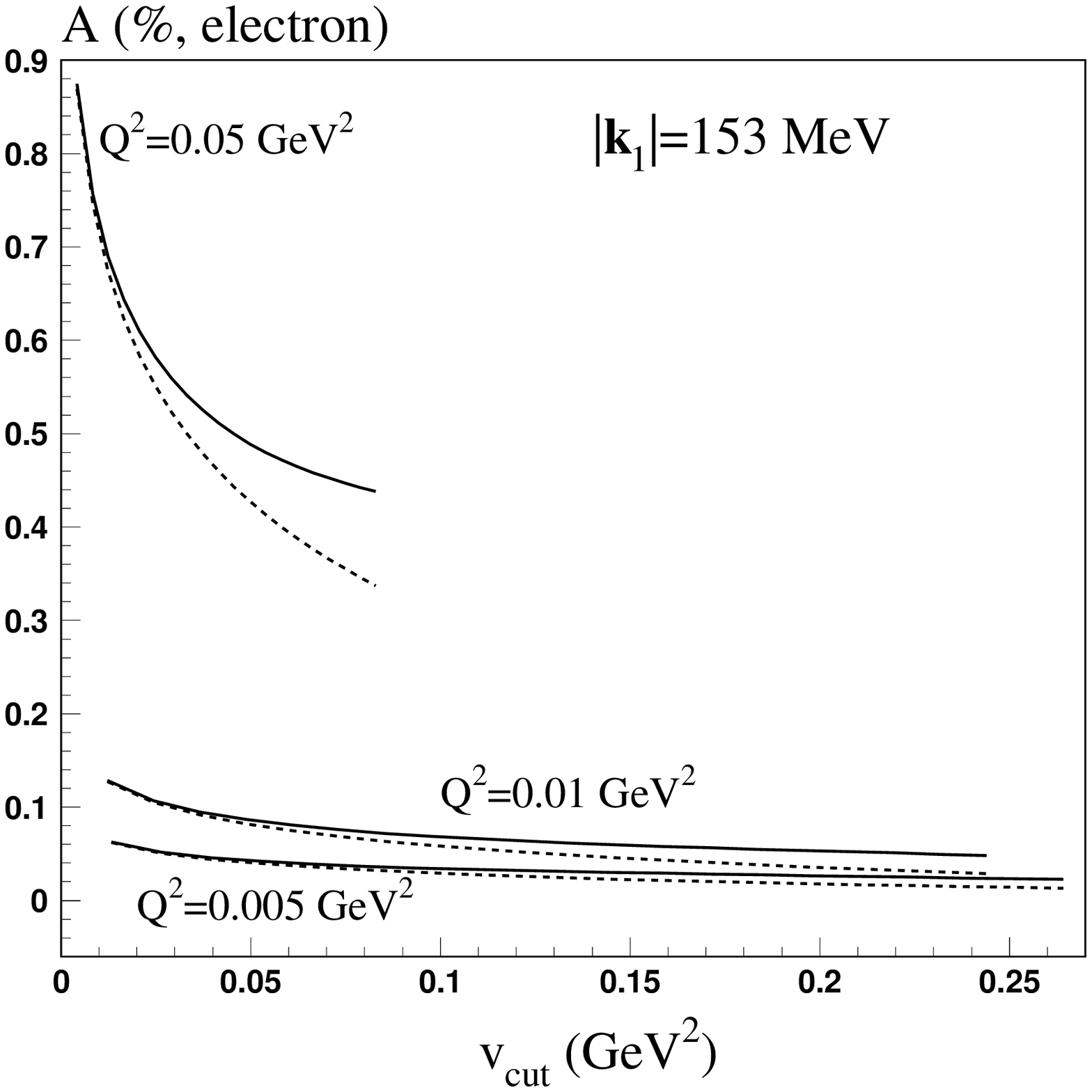}
\hspace*{-6mm}
\includegraphics[width=70mm,height=70mm]{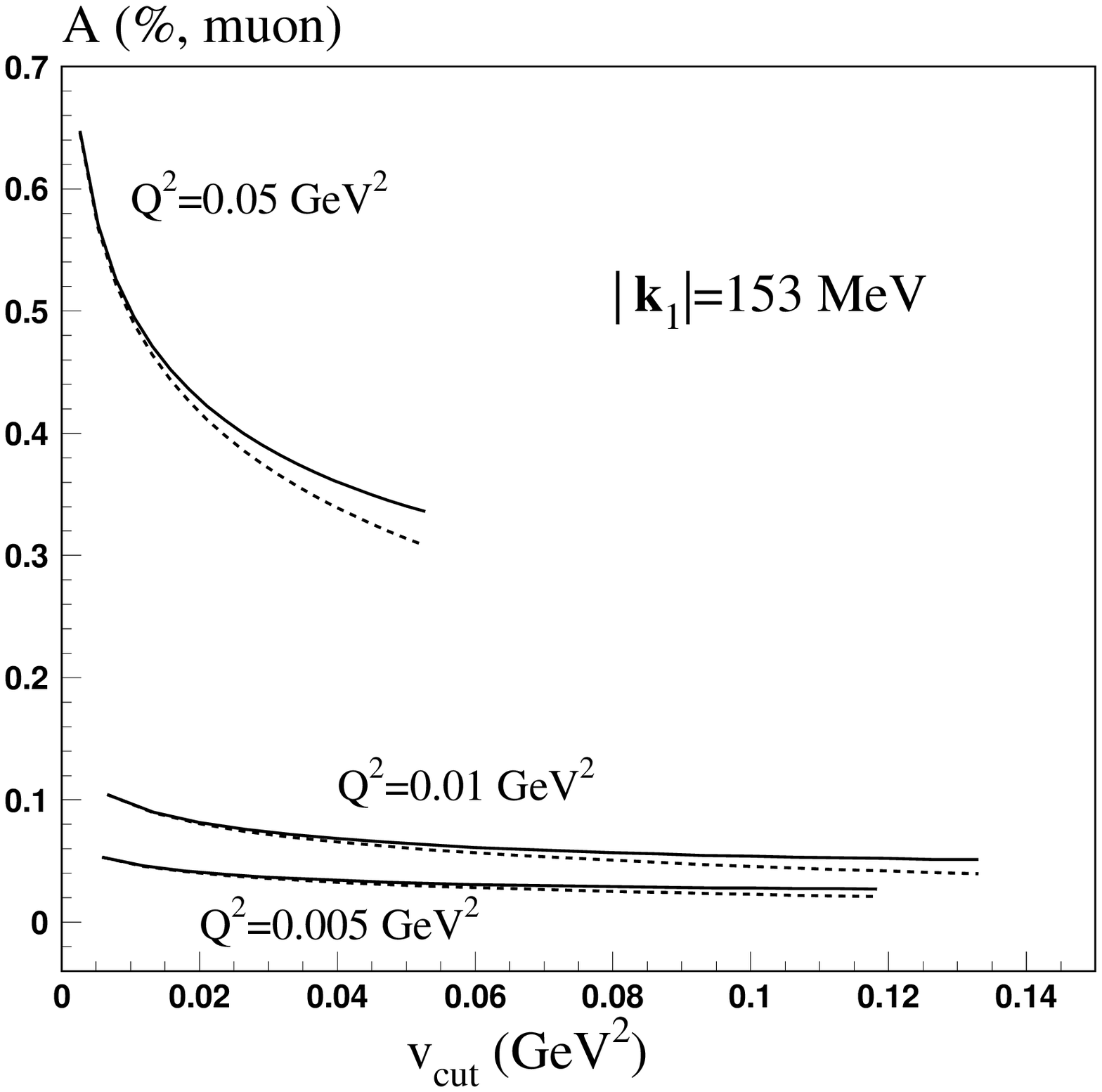}
\\[-9mm]
\includegraphics[width=70mm,height=70mm]{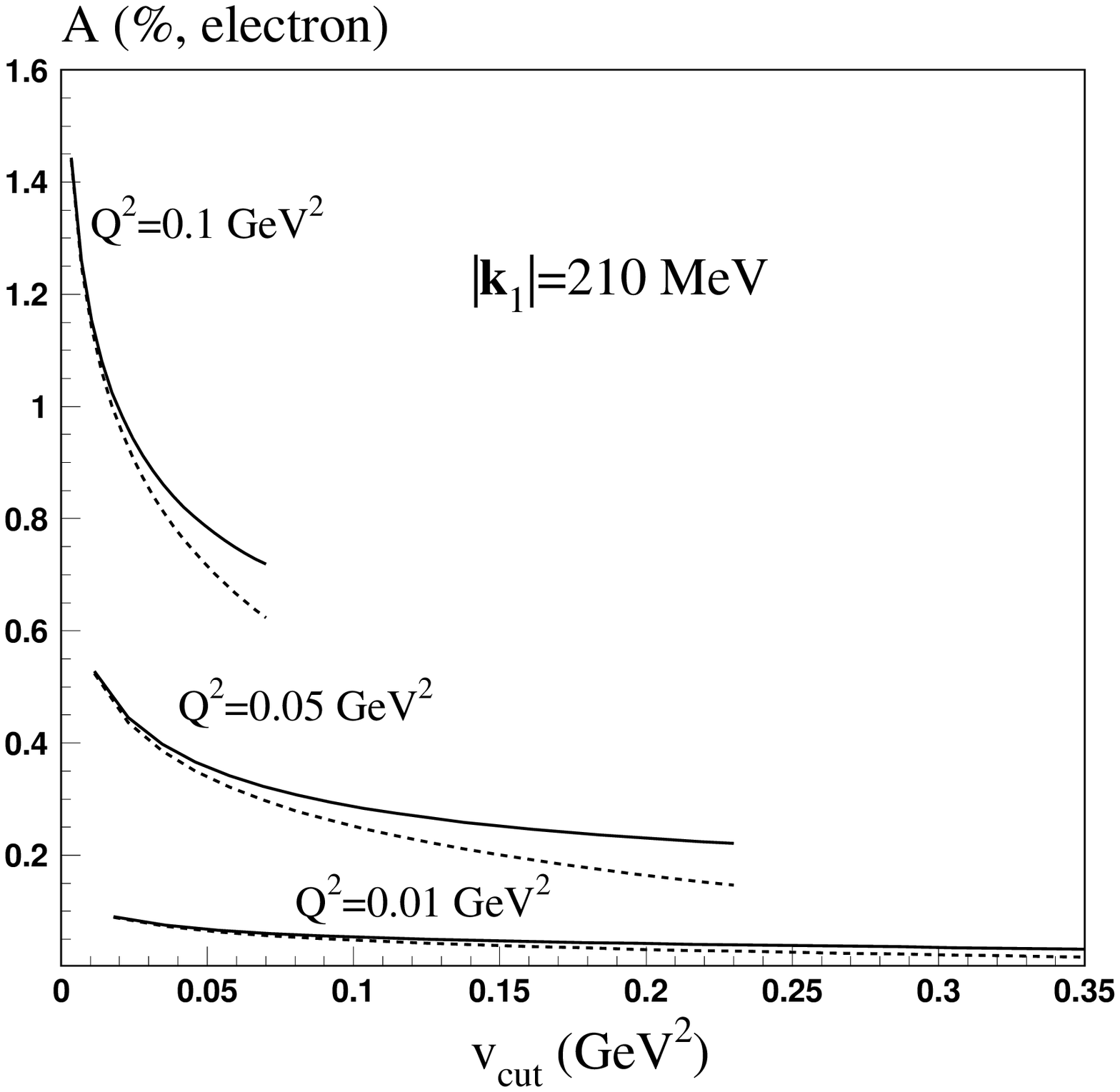}
\hspace*{-6mm}
\includegraphics[width=70mm,height=70mm]{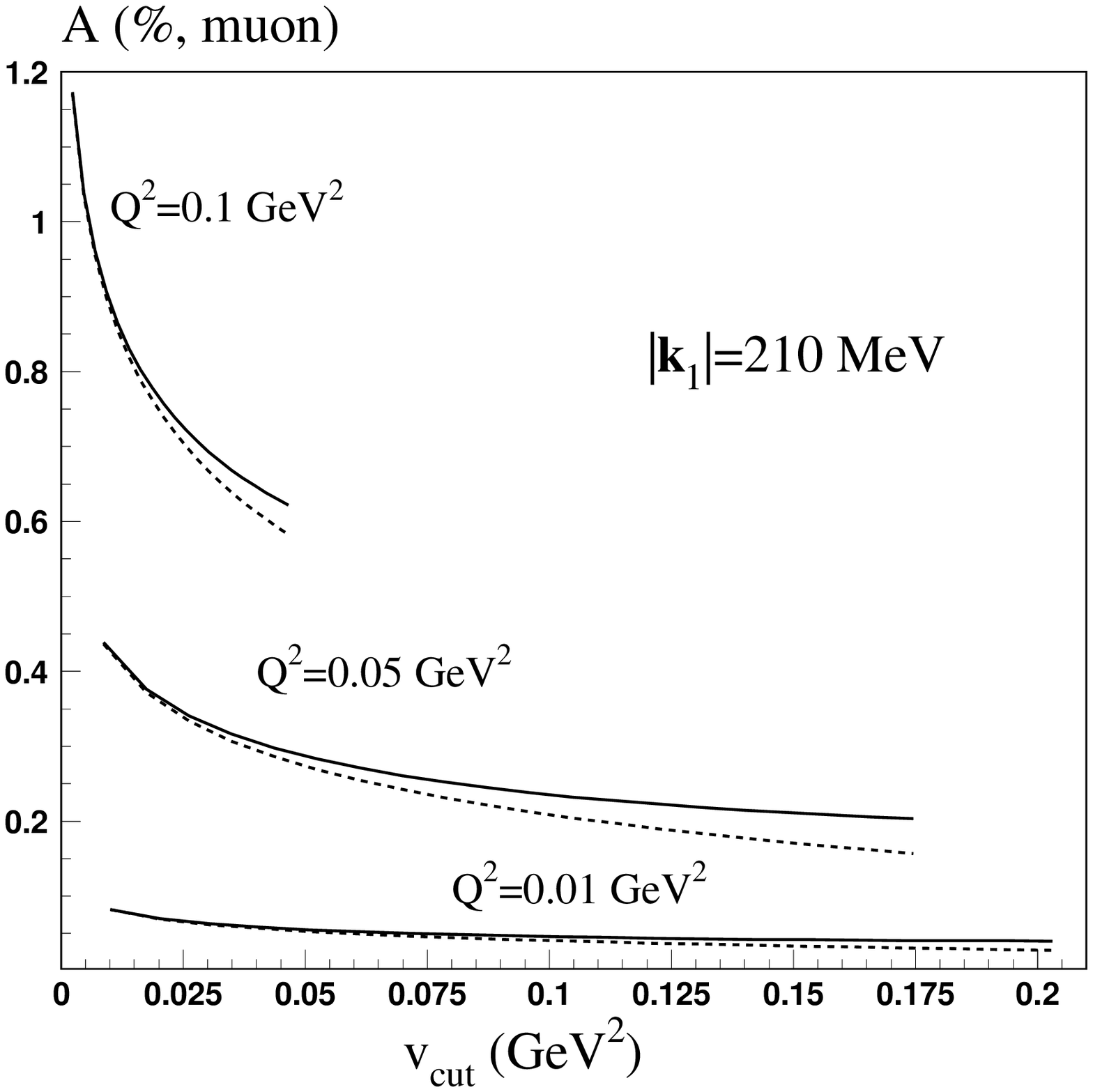}
\\[-4mm]
\caption{Charge asymmetry defined in (\ref{casy}) vs value of the inelasticity cut
for elastic $e^\mp p$ and $\mu^\mp p$ scattering with the beam momenta 115 MeV, 153 MeV and 210 MeV. 
The dashed lines correspond to the soft photon approximation \cite{Koshchii}. The solid lines include the hard photon emission.
}  
\label{fig3}
\end{figure*}
\begin{figure}[t]
\vspace*{-4mm}
\includegraphics[width=75mm,height=75mm]{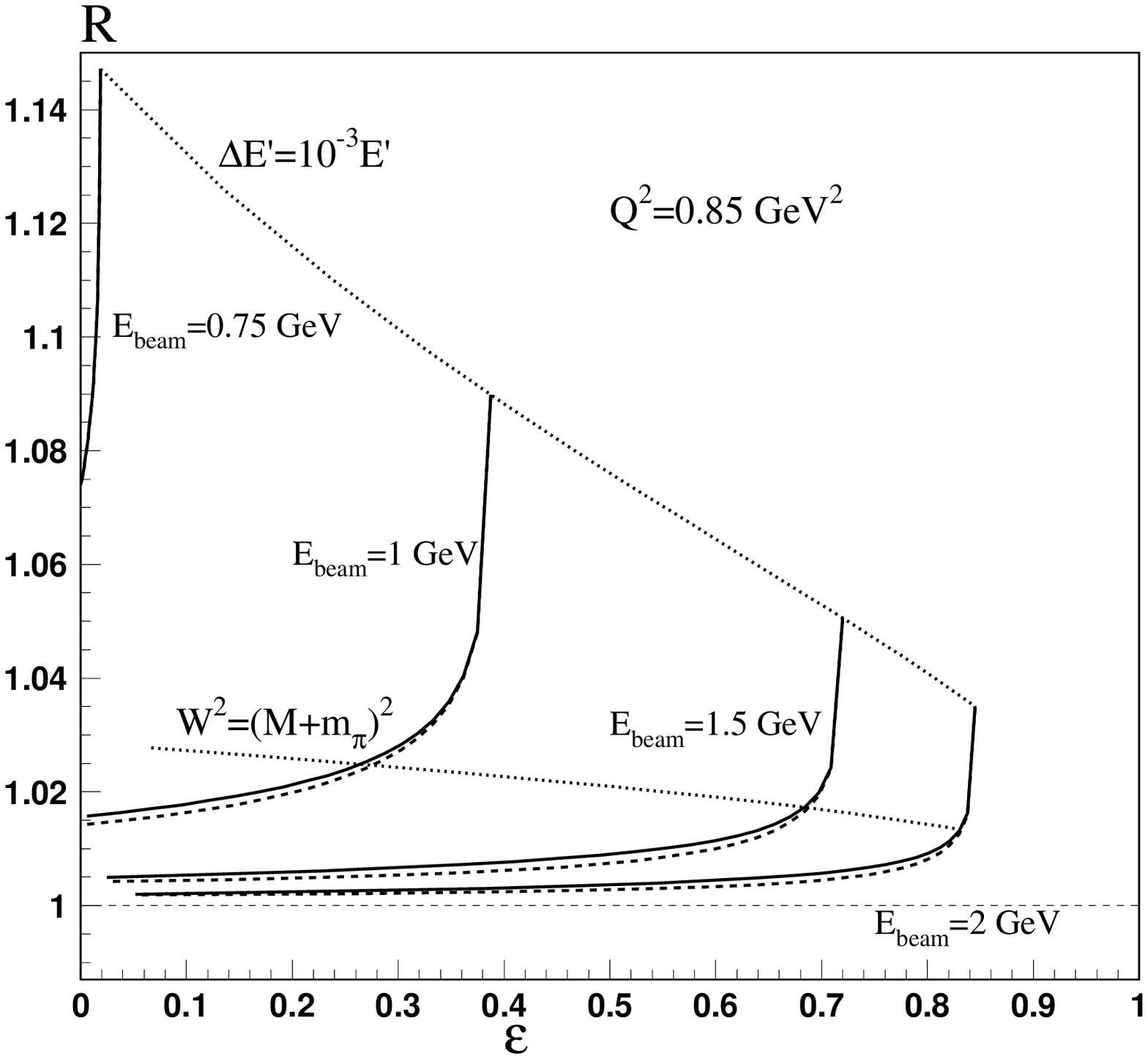}
\\
\vspace*{-8mm}
\includegraphics[width=75mm,height=75mm]{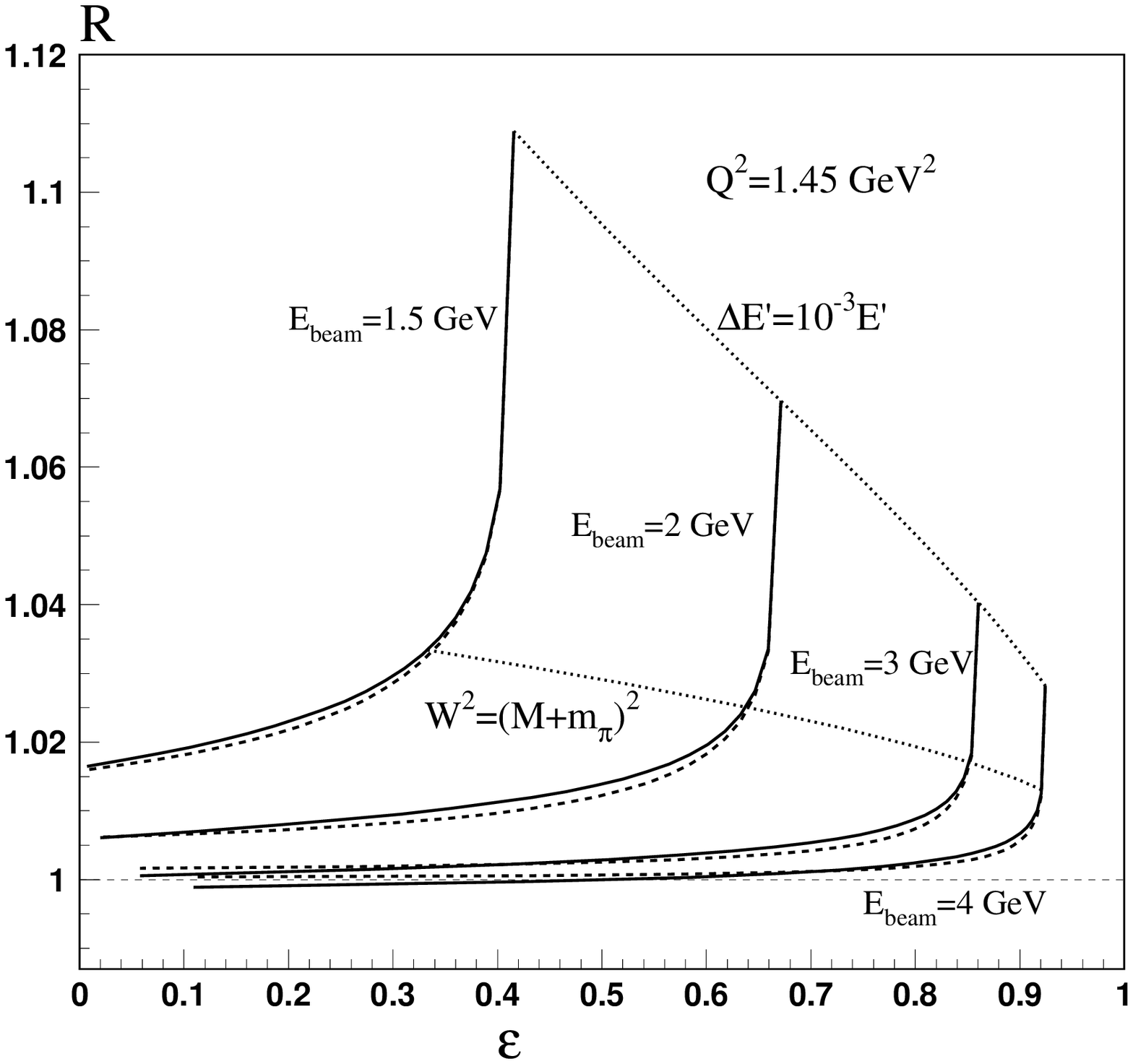}
\caption{
Ratio of $e^+p/e^-p$ cross sections defined in (\ref{rat}) as a function of $\varepsilon$ at $Q^2=0.85$ GeV$^2$ and $Q^2=1.45$ GeV$^2$.
Solid lines include the hard photon emission while dashed ones correspond to the soft photon approximation \cite{Koshchii} for the different electron beam energies $E_{beam}$. 
An upper dotted lines present the minimum detector resolution over scattering electron energy $\Delta E^\prime=10^{-3}E^\prime$. The lowest dotted lines correspond
to the pion production threshold.}
\label{fig4}
\end{figure}

\section{Numerical results}
The dependence of the charge asymmetry   
\begin{eqnarray}
A=\frac{d\sigma^{+}_{odd}/dQ^2-d\sigma^{-}_{odd}/dQ^2}{d\sigma_B/dQ^2},
\label{casy}
\end{eqnarray}
on the value of the upper integrated limit over inelasticity $v_{cut}$ for different lepton beams and $Q^2$ is presented in Fig.~\ref{fig3} at MUSE kinematic conditions \cite{MUSE}. The dashed lines correspond to the contribution only the soft photon approximation
calculated in \cite{Koshchii} while the solid lines include the hard photon emission presented in the given work. From these plots one can see that at the fixed lepton momentum the asymmetry decreases more rapidly without hard photon contribution with growing $v_{cut}$. It can be also seen that the magnitude of the 
asymmetry is reduced with decreasing of
$Q^2$.  Moreover, the value of this asymmetry is higher for the lighter lepton.

Another important quantity is the ratio of $e^+p/e^-p$ cross sections that can be defined as
\begin{eqnarray}
R=\frac{d\sigma_B/dQ^2+d\sigma^{+}_{odd}/dQ^2}{d\sigma_B/dQ^2+d\sigma^{-}_{odd}/dQ^2}.
\label{rat}
\end{eqnarray}

The dependence of this quantity on the virtual photon polarization $\varepsilon$ 
at JLab kinematic conditions is shown on Fig.~\ref{fig4}. Since beyond the ultrarelativistic approximation $\varepsilon$ for the bremsstrahlung process depends on $v_{cut}$ as  
\begin{eqnarray}
\varepsilon=\Biggl(1+2(1+\tau )\frac{M^2(Q^2-2m^2)}{S(X-v_{cut})-M^2Q^2}\Biggr)^{-1}
\label{eps}
\end{eqnarray}
for the fixed electron beam energies (solid lines) the small $\varepsilon$
corresponds to the hard photon contribution where the ratio (\ref{rat}) is closer to one. 
Like to previous plots the soft photon approximation presented by dashed lines while the solid lines contain the hard photon emission.
Similar to the experimental observation \cite{Jlab,Jlab1}, the cross section ratio
with the soft photon emission corresponding the minimum detector resolution over scattering electron energy $\Delta E^\prime=10^{-3}E^\prime$ (upper dashed lines) decreases with growing $\varepsilon$.
The lowest dashed lines correspond the pion threshold when the quantity $W^2=(p_1+q)^2$ reaches a value $(M+m_{\pi})^2$ and 
together with the real photon one undetected  pion could to be produced.

\section{Conclusion}
The contribution of the hard photon emission to the charge asymmetry in lepton-proton scattering 
was estimated for the first time beyond the ultrarelativistic limit, while keeping lepton mass during 
the entire calculation.

Only  two assumptions were used in the calculation: I) We did not consider excitations of the intermediated proton and used a standard fermionic propagator for it; II) The on-shell proton vertex with the Dirac and Pauli form factors were used in the off-shell region.

Numerical results shown that 
at the fixed lepton momentum the charge asymmetry decreases 
with growing  both the energy of the unobserved photon and the transferred momentum square $Q^2$.
This asymmetry is sensitive to the lepton mass: its value
is higher for the lighter lepton.

The next planned step consists in implementation of the obtained results into Monte-Carlo generator ELRADGEN \cite{ELRADGEN1,ELRADGEN2} for 
simulation of hard photon emission
in future experiments.

\bibliography{chasum}
\end{document}